\documentclass[aps, prl, twocolumn, english, notitlepage
]{revtex4-2}

\usepackage{graphicx}% Include figure files
\usepackage{dcolumn}% Align table columns on decimal point
\usepackage{bm}% bold math
\usepackage[colorlinks=true, linkcolor=blue, citecolor=blue, urlcolor=blue]{hyperref}% add hypertext capabilities
\usepackage{mhchem}
\usepackage{amsmath}
\usepackage{physics}
\usepackage{makecell}

\usepackage{pdfpages} % include pdfs
\usepackage{pgffor} % for loops

\makeatletter
\AtBeginDocument{\let\LS@rot\@undefined}
\makeatother

\def\supplementfilename{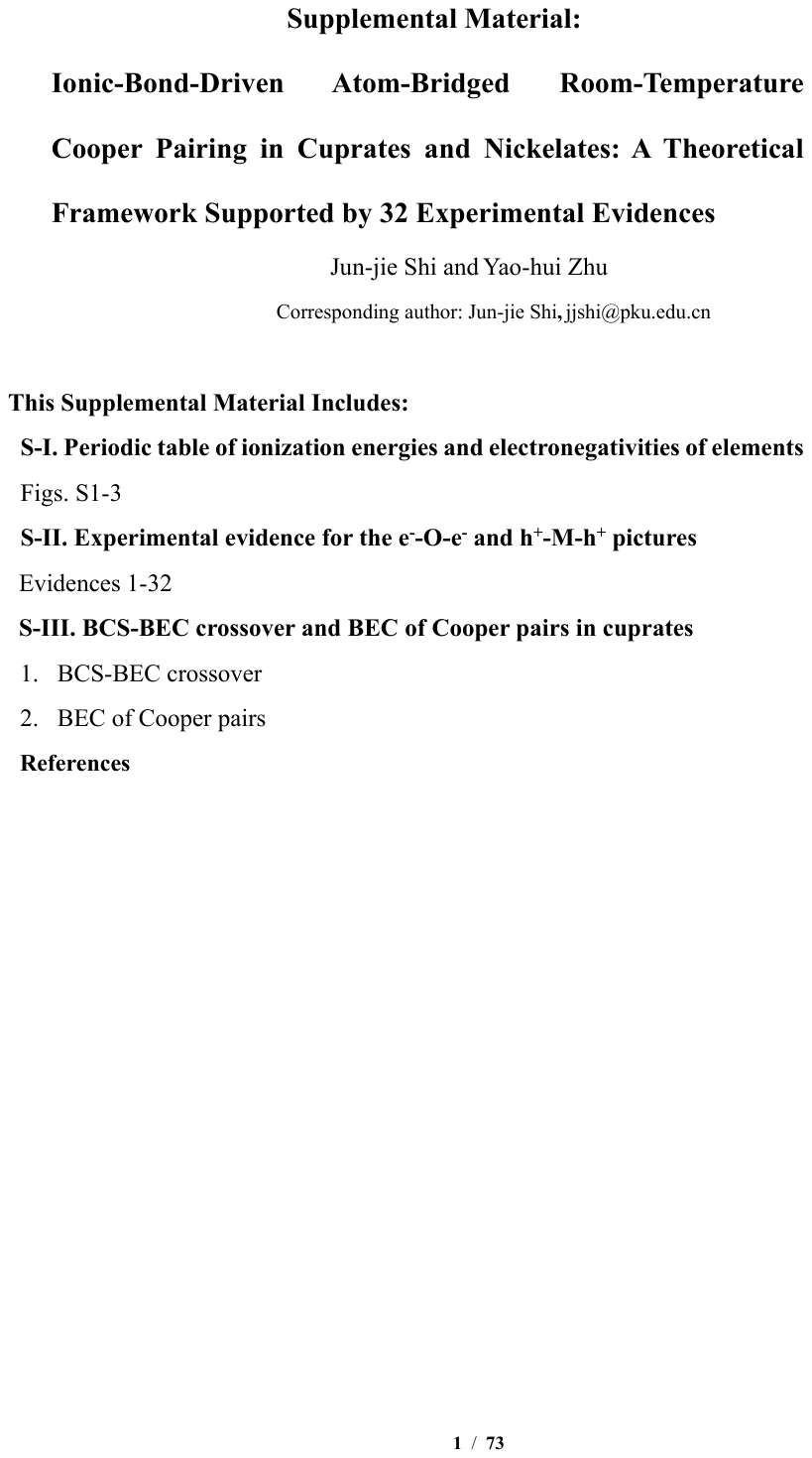}

\pdfximage{\supplementfilename}
\def\numbersupplementpages{\the\pdflastximagepages}

\newif\ifarXiv
\arXivtrue 
\begin{document}

\preprint{APS/123-QED}

\title{Ionic-Bond-Driven Atom-Bridged Room-Temperature Cooper Pairing in Cuprates and Nickelates: A Theoretical Framework Supported by 32 Experimental Evidences}% Force line breaks with \\

\author{Jun-jie Shi\textsuperscript{1,$^*$} and Yao-hui Zhu\textsuperscript{2}}

\affiliation{\textsuperscript{1}State Key Laboratory for Artificial Microstructures and Mesoscopic Physics, School of Physics, Peking University Yangtze Delta Institute of Optoelectronics, Peking University, Beijing 100871, China}
\affiliation{\textsuperscript{2}Physics Department, Beijing Technology and Business University, Beijing 100048, China}

\date{\today}% It is always \today, today,

\begin{abstract}
Unlike ordinary conductors and semiconductors, which conduct electricity through individual electrons, superconductors usually conduct electricity through pairs of electrons, known as Cooper pairs. Even after 4~decades of intense study, no one knows what holds electrons together in complex high-$T_c$ cuprates. Here, targeting the critical challenge of the pairing mechanism behind high-$T_c$ superconductivity in oxides and considering the dominance of eV-scale ionic bonding, affinity of O$^-$ (1.46 eV) and O$^{2-}$ (-8.08 eV) and large two-electron ionization energy ($\sim$15-28 eV) of metal atoms, we propose a groundbreaking idea of electron e$^-$ (hole h$^+$) pairing bridged by oxygen O (metal M) atoms, i.e., the ionic-bond-driven \textbf{e$^-$-O-e$^-$} (\textbf{h$^+$-M-h$^+$}) itinerant Cooper pairing formed at pseudogap temperature $T^*$>$T_c$, by following the principle of “tracing electron footprints to explore pairing mechanisms” and by standing on the solid foundation of the chemical-bond$\rightarrow$structure$\rightarrow$property relationship. It is applicable to cuprates, nickelates, iron-based and other new potential ionic superconductors. Its correctness and universality are confirmed by 32 diverse experimental evidences, especially, the STM image in the CuO$_2$ plane combining with the small Cooper-pair size as an ironclad proof. Any other sub-eV and covalent-binding pairing mechanisms would be doubtful. Our findings, which provide the missing link between ionic bonding and superconductivity, resolve a 40-year puzzle and validate the feasibility of room-temperature carrier-pairing in ionic-bonded superconductors. We further create a new theoretical framework rooted in our universal \textbf{e$^-$-O-e$^-$} (\textbf{h$^+$-M-h$^+$}) picture with the strongest pairing strength and Bose-Einstein condensation, which opens a new avenue for understanding high-$T_c$ mechanism and brings the dream of room-temperature superconductivity one step closer.
\end{abstract}

\maketitle

Since the discovery of high-$T_c$ superconductivity in cuprates in 1986~\cite{Bednorz1986}, nickelates have recently emerged as another family exhibiting high-$T_c$ behavior~\cite{Wang2024,WangNingning2024}. A common feature among these oxides is their oxygen-rich composition. Unfortunately, the microscopic mechanism underlying their unconventional superconductivity remains elusive—a significant challenge called as the scientific Tower of Babel by Nobel Prize laureate P.W. Anderson, which will lead to rewriting of the condensed matter textbooks believed by Nobel Prize winner J.R. Schrieffer~\cite{Anderson1991}.

It is widely accepted that superconductivity arises from two steps: the formation of Cooper pairs and their condensation. To uncover the carrier-pairing mechanism in cuprates and nickelates, such as YBa$_2$Cu$_3$O$_{6+x}$, $T_c$$\sim$93~K~\cite{Mourachkine2002}, RNiO$_2$ (R=La,Pr,Nd), $T_c$$\leq$23~K~\cite{Wang2024} and La$_2$PrNi$_2$O$_7$ with $T_c^{zero}$=60~K at 18-20 GPa~\cite{WangNingning2024}, we begin by analyzing their chemical compositions. Oxygen is the sole non-metal element in these superconductors. Metal atoms, with relatively low first ionization energies about 5-8 eV (Figs.~S1-3 in Supplemental Material (SM)~\cite{Supplemental})~\cite{Kittel2005,Landau1977}, readily donate their outer electrons, which contribute to electrical conduction. In contrast, oxygen, being highly  electronegative (second only to fluorine), naturally attracts electrons, thereby acting as a natural electron gathering center. Consequently, oxygen plays a critical role in the unconventional superconductivity observed in electron-doped cuprates and nickelates, as well as in their microscopic electron-pairing mechanism. Conversely, in hole-doped high-$T_c$ oxides, metal atoms function as the centers for hole accumulation and pairing.

As the eighth element in the periodic table, oxygen possesses a nucleus that strongly binds its outer electrons. Despite the significant Coulomb repulsion between electrons, the oxygen nucleus, with strong attraction to electrons, robustly holds its eight electrons in the  $1s^22s^2p^4$ configuration. Moreover, oxygen can accept two additional electrons to make its $2p$ orbital fully occupied, forming a closed electron shell of $1s^22s^2p^6$ (O$^{2-}$). However, in the case of isolated atoms, this process is completely forbidden, indicating O$^{2-}$'s instability (see End Matter (EM))~\cite{Atkins2010}. On the contrary, in ionic oxides, although the O$^-$ anion is easily formed to repel electrons, the strong ionic bonding and strong attraction of oxygen nucleus to electrons will drive the O$^-$ anion to capture 0-1 additional electron to form higher valence anion O$^{x-}$ (1$<$$x$$\leq$2).

\begin{figure}
    \centering  \includegraphics[width=\linewidth]{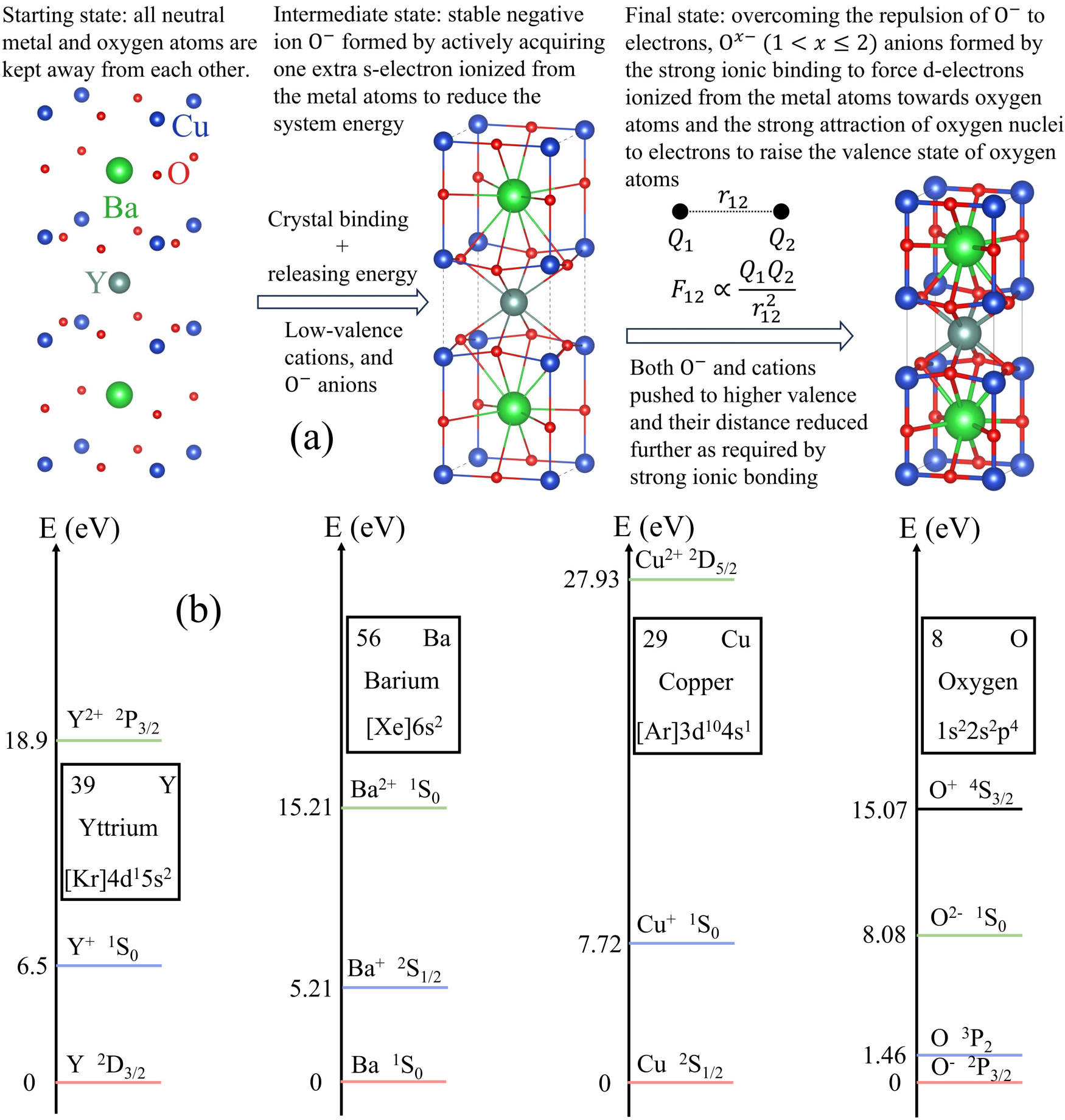}
    \caption{Formation process of ionic oxides with low-valent ions: carrier migration and gathering. (a) We take high-$T_c$ superconductor \ce{YBa2Cu3O7} (YBCO) as an example to describe the crystal formation process in order to explore the pairing mechanism of electrons by tracing their footprints. (b) Energy levels of neutral metal atoms Y, Ba, Cu and their respective cations \ce{Y^{\textrm{n}+}}, \ce{Ba^{\textrm{n}+}}, \ce{Cu^{\textrm{n}+}} (n=1,2), and the oxygen atom O, its anions (\ce{O^-}, \ce{O^{2-}}) and cation (\ce{O+})~\cite{Atkins2010,Kittel2005}. Here, the levels of Y, Ba, Cu and O$^-$ are set to zero, respectively. We find that the strength of ionic bonds is in the magnitude of eV, about 2 orders of magnitude larger than antiferromagnetic coupling in YBCO ($J_c$=12~meV, $J_{ab}$=120~meV)~\cite{Mourachkine2002} and electron-phonon interaction in cuprates ($\sim$40-80~meV)~\cite{Yan2023,Lanzara2001}. The large two-electron ionization energy ($\sim$15-28~eV) of metal atoms and the electron affinity (-8.08~eV) of \ce{O^{2-}} ensure that ions can only exist in the low-valent states, which is a key for us to solve the high-$T_c$ mechanism in cuprates and nickelates.}
    \label{fig:1}
\end{figure}

\begin{figure}
    \centering  \includegraphics[width=\linewidth]{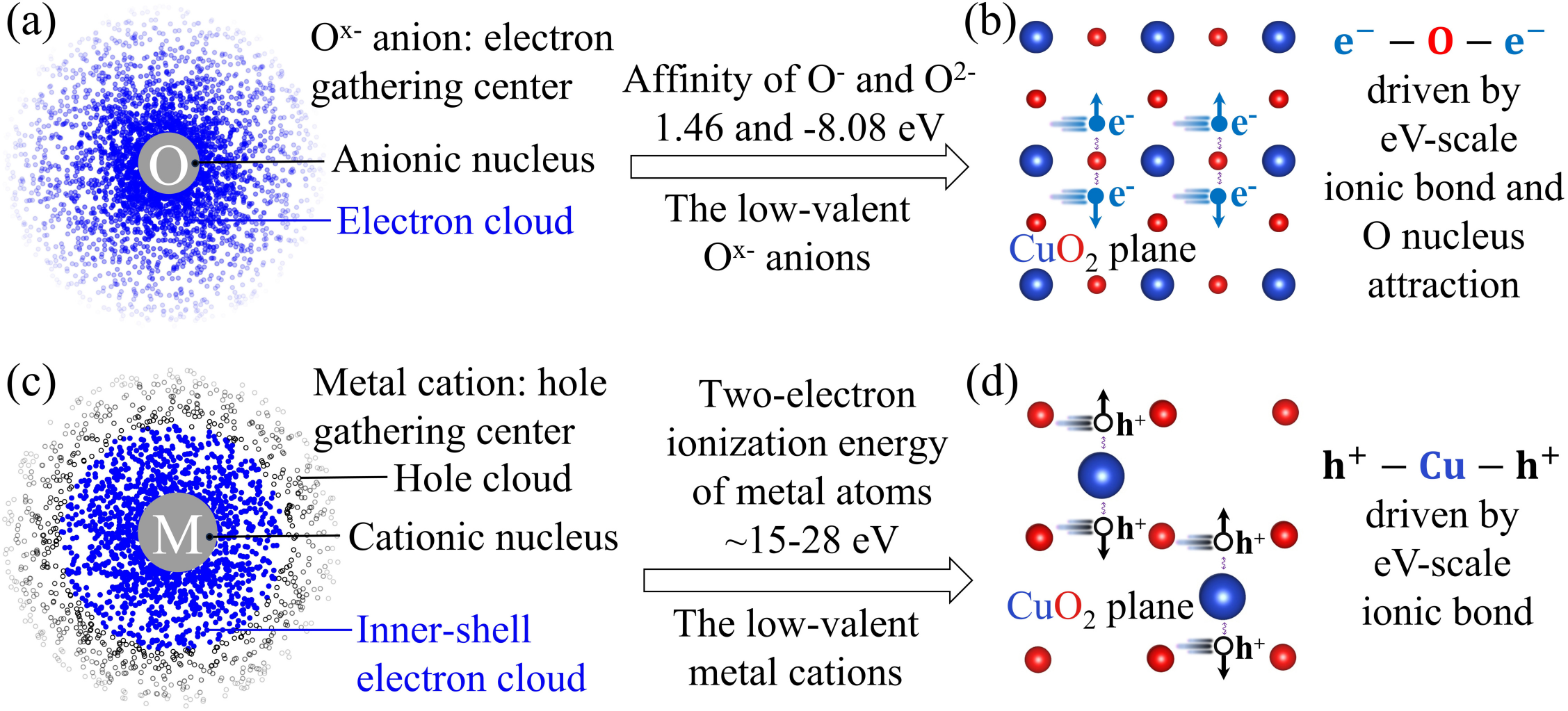}
    \caption{From the electron (hole) clouds to the atom-bridged Cooper pairing picture \textbf{\ce{e-}-O-\ce{e-}} (\textbf{\ce{h+}-Cu-\ce{h+}}). As the most basic constituent ions in ionic oxides, the O$^{x-}$ anion surrounded by the electron cloud and metal (M) cation together with its inner-shell electrons surrounded by the hole cloud due to the strong ionic bonds are shown in (a) and (c), respectively. (b) Due to the low-valent ions, the conduction electrons, i.e., the $d$-electrons (the origin and solid evidence of $d$-wave superconductivity) donated by metal atoms, which are not captured by oxygen atoms, are forced to gather towards oxygen atoms by the eV-scale ionic bonding and the strong attraction of oxygen nucleus to them, forming itinerant superconducting $d$-wave electron pairs above $T_c$~\cite{Mourachkine2002,Xiang2022,Gomes2007,Kondo2011}. It is this O-bridged electron pair \textbf{\ce{e^-}-O-\ce{e^-}}, featured with large pseudogap and small pair size comparable to the lattice constant, that dominates the high-$T_c$ superconductivity of oxides. The two pairs of electrons are bridged by two respective oxygen anions here. (d) Same as in (b), but for the Cu-bridged $d$-wave hole pairs \textbf{\ce{h+}-Cu-\ce{h+}}. The key data of the affinity of O$^-$ and O$^{2-}$ and the large two-electron ionization energy of metal atoms are taken from textbooks~\cite{Atkins2010,Kittel2005}, respectively. The spatial symmetry of wavefunctions necessitates that two electrons (holes) approach each other to form Cooper pairs with opposite spins~\cite{Sakurai1994}.}
    \label{fig:2}
\end{figure}

\textit{Low-valent ions: carrier migration and gathering}---It is well-known that, according to the Coulomb’s law, the strength of ionic bonds depends on the distance between ions and the amount of charge carried by them. Taking cuprates as an example, when initially isolated neutral atoms coalesce to form an ionic crystal, the energy released by the system first ionizes the outermost $s$-electrons of the metal atoms due to their low first ionization energy (Figs.~S1-3~\cite{Supplemental}). The neutral O atoms then capture these electrons to form \ce{O-} anions, further lowering the system’s energy~\cite{Atkins2010}. Consequently, the system becomes populated with low-valent metal cations and \ce{O-} anions. To achieve even greater energy reduction, the system promotes the ionization of $d$-electrons from the low-valent cations, raising their oxidation states. These ionized $d$-electrons are subsequently drawn to the \ce{O-} anions, increasing their valence states while reducing the interionic distances. This process ultimately results in a stable structure with the lowest total energy (Fig.~\ref{fig:1}(a)).

\begin{figure}
    \centering  \includegraphics[width=8 cm]{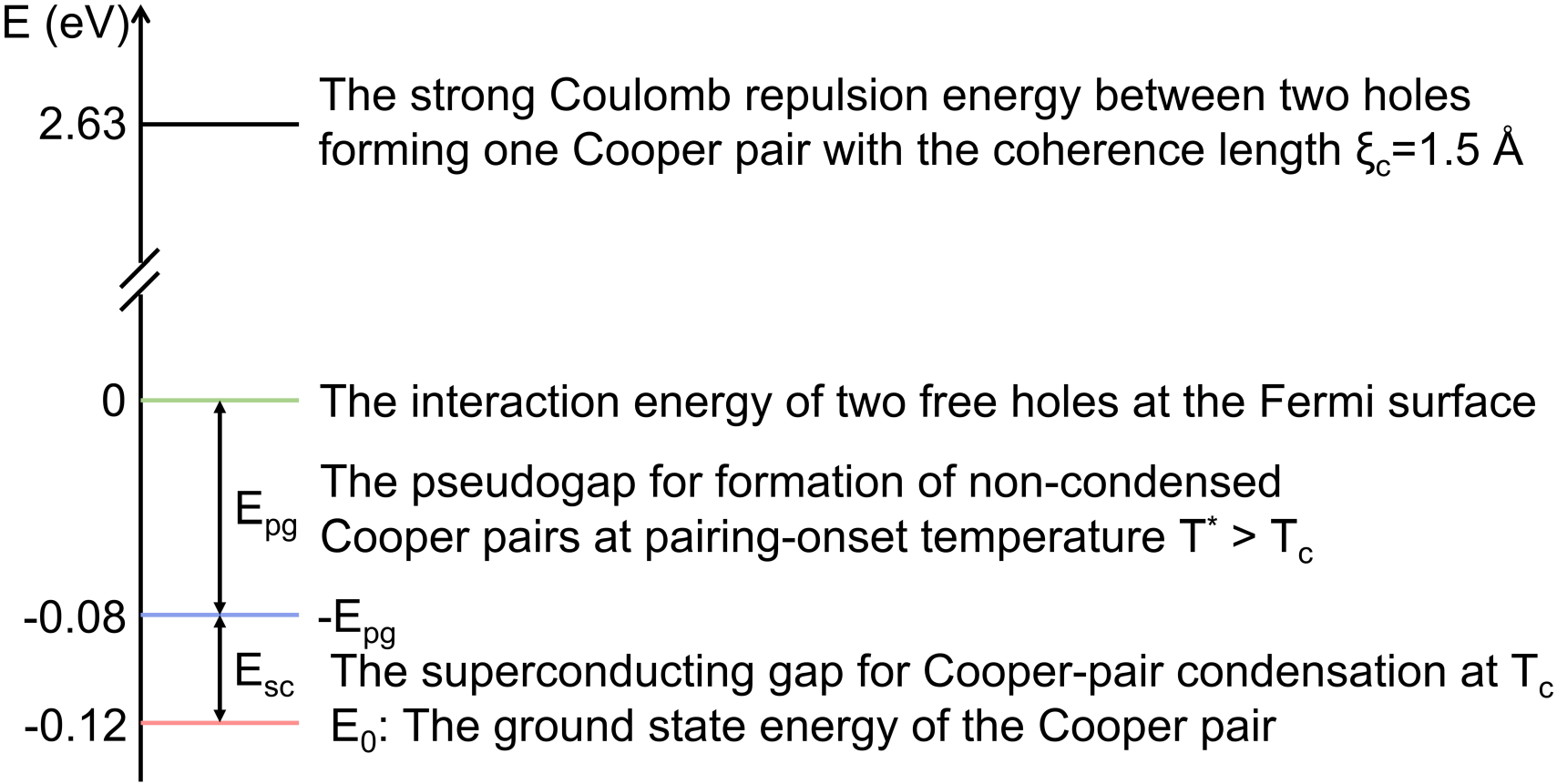}
    \caption{Energy scales in high-$T_c$ cuprates from two repulsive holes to the hole-pair formation and condensation into the superconducting state. Here, we set the interaction energy between two free holes to zero. Considering the short coherence length of $\xi_c$=1-2.5~$\textrm{\AA}$ of the hole Cooper pairs~\cite{Mourachkine2002}, as a lower limit, we estimate the corresponding Coulomb repulsion energy about 6.08-0.67~eV, confirmed by Refs.~\cite{Anderson1959,Derriche2025,Alexandrov2013,Ohta1991}, between the two holes forming a Cooper pair under the strongest screening with the maximum screening electron concentration of $1 \times 10^{21}$~cm$^{-3}$ at the same level as the hole concentration, in which the Thomas-Fermi screening is included with the screening length of 1.16~$\textrm{\AA}$ as the major screening mechanism and the dielectric screening is minor due to the smaller coherence length than the lattice constant ($\sim$4-5~$\textrm{\AA}$). It is evident that only the eV-scale ionic bonds can overcome the strong Coulomb repulsion between two holes, and confine hole pairs within the \ce{CuO2} plane (Evidence~2~\cite{Supplemental}), which can be regarded as an ironclad proof of \textbf{\ce{h+}-Cu-\ce{h+}} picture (Fig.~\ref{fig:2}(d)). According to experiments of high-$T_c$ cuprates~\cite{Hüfner2008}, we adopt a pseudogap of approximately 0.08~eV and a superconducting gap of about 0.04~eV, smaller than the pseudogap, for the optimally doped \ce{Bi2Sr2CaCu2O_{8+\delta}} (Bi2212) as references for the pseudogap and superconducting gap, corresponding to the formation and condensation of Cooper pairs (see Evidences:~7-12~\cite{Supplemental}), respectively‌. For other cuprates with different doping concentrations, their pseudogap and superconducting gap exhibit certain fluctuations around these two reference values‌.}
    \label{fig:3}
\end{figure}

Figure~\ref{fig:1}(b) illustrates that the strength of ionic bonds is on the order of eV, exceeding the antiferromagnetic or electron-phonon coupling in cuprates by approximately two orders of magnitude~\cite{Mourachkine2002,Yan2023,Lanzara2001}. This significant energy difference casts doubt on the viability of magnetic or electron-phonon coupling-based pairing mechanisms or other pairing patterns within the sub-eV scale due to their weak interaction (Fig.~\ref{fig:3}). It is the strong ionic bonding requires that oxygen atoms exist as \ce{O^{2-}} anions as much as possible. As a result, the repulsive effect of \ce{O^{-}} to electrons is ultimately suppressed by the interplay of strong ionic bonding and the intrinsic attraction of the oxygen nucleus to electrons. This leads to an eV-scale ``net attraction'' of \ce{O^{-}} to electrons, which facilitates the accumulation of $d$-electrons donated by metal atoms around \ce{O^{-}} anions to reduce the total energy of the crystal and to maintain structural stability. Similarly, holes with $d$-wave symmetry cluster around the metal cations because of the driving of strong ionic bonding.

The competition between the strong ionic bonding, the attraction of oxygen nucleus, and the repulsion of O$^-$ to electrons results in a balanced valence state of O$^{x-}$ (1$<$$x$$\leq$2) in ionic oxides. Notably, in \ce{YBa2Cu3O7}, the average valence states of O$^{x-}$ anions and Cu$^{y+}$ cations are low, calculated as -1.69 and +1.62~\cite{Krakauer1988}, respectively. This finding confirms that in ionic oxides, despite their negative charge, O$^{x-}$ anions with a low-valent state $x$$<$$2$ exhibit a strong net attraction toward electrons rather than repulsion. This net attraction arises from the combined effects of ionic bonding, which drives electrons toward O atoms, and the strong nuclear attraction exerted by oxygen. Consequently, neutral O atoms and O$^-$ anions naturally serve as electron-gathering centers and potential electron-pairing mediators in high-$T_c$ oxides.

Figure~\ref{fig:2} schematically depicts the spatial distribution of electron and hole clouds surrounding O$^{x-}$ anions and metal cations, respectively. Moreover, the important connection between electron (hole) cloud and the corresponding electron (hole) Cooper pairing is established through the solid physical foundation of the affinity of O$^-$ (1.46~eV) and O$^{2-}$ (-8.08~eV)~\cite{Atkins2010} and large two-electron ionization energy ($\sim$15-28~eV) of metal atoms~\cite{Kittel2005}. \textit{It is the electrons (holes) located at the indistinct boundary of the atom, furthest from the atomic nucleus and not fully captured by it (the low-valent ion), that undergo pairing under the driving of eV-scale ionic bonding, forming itinerant superconducting Cooper pairs.}

Figure~\ref{fig:3} demonstrates that to overcome the strong Coulomb repulsion between two holes to form the superconducting Cooper pair, the external energy driving their pairing must be quantitatively in the eV scale. Evidently, only the strong ionic bonding and the strong nuclear attraction to electrons can fulfill this critical role. Any other sub-eV pairing mechanisms would be insufficient to overcome the strong Coulomb repulsion between two holes very close to each other, making it impossible to form the superconducting Cooper pair. Furthermore, in addition to driving hole Cooper pairing, the eV-scale ionic bonds can also overcome the carrier thermal motion energy of about 26~meV at room temperature, ensuring that the Cooper pair is not destroyed by thermal motion at room temperature. This provides a solid foundation for achieving room-temperature superconductivity. Figure~\ref{fig:3} also clarifies the physical origins of the pseudogap and superconducting gap~\cite{Hüfner2008}, i.e., the formation and condensation of Cooper pairs (Evidences:~7-12~\cite{Supplemental}).

\textit{Atom-bridged room-temperature Cooper pairing via ionic bonds}---Based on the eV-scale ``net attraction'' of O atoms and O$^-$ anions to electrons and the requirements of the ionic bonds for electron transfer and gathering during the crystal formation as analyzed above, we know that the high-$T_c$ ionic oxides have a $d$-wave symmetric electron-pairing image due to the strong ionic bonds driving the itinerant $d$-electrons to gather around oxygen atoms~\cite{Xiang2022}: \textbf{\ce{e^-}-O-\ce{e^-}} (Fig.~\ref{fig:2}(b)). Generally, the ground ($E_0$) and excited ($E_n$, $n$=1,2,3,...) state wavefunctions of the \textbf{\ce{e^-}-O-\ce{e^-}} Cooper pair can be written as~\cite{Sakurai1994},
\begin{widetext}
\begin{equation}
\label{eqn:1}
\begin{aligned}
    \Phi_n (\boldsymbol{r}_1, s_{1z};\, \boldsymbol{r}_2, s_{2z})
= \psi_{nS} (\boldsymbol{r}_1, \boldsymbol{r}_2)\, \chi_A (s_{1z}, s_{2z})
\xrightarrow[\boldsymbol{r} = \boldsymbol{r}_1 - \boldsymbol{r}_2]{\boldsymbol{R} = (\boldsymbol{r}_1 + \boldsymbol{r}_2) /2}
\Theta_n (\boldsymbol{R})\, \phi_{nS}(\boldsymbol{r})\, \chi_A(s_{1z}, s_{2z}),
    \end{aligned}
\end{equation}
\end{widetext}
where $\chi_A$ is the well-known antisymmetric spin wavefunction of the two-electron system, $\Theta_n (\bm{R})$ is the wavefunction of the center-of-mass motion with characteristics of the tight-binding approximate wavefunction~\cite{Kittel2005}, and the symmetric spatial wavefunction $\phi_{nS}(\boldsymbol{r}$) is the eigenfunction of the orbital angular momentum $\bm{\hat{L}} = -i \hbar \bm{r} \times\nabla$,
\begin{equation}
\label{eqn:2}
\begin{aligned}
    \bm{\hat{L}}^2\, \phi_{nS}(\bm{r}) = l(l + 1)\, \hbar^2\, \phi_{nS}(\bm{r}), (l=2),
    \end{aligned}
\end{equation}
\begin{equation}
\label{eqn:3}
\begin{aligned}
    \phi_{nS}(\bm{r}) = \phi_{nS}(-\bm{r}).
    \end{aligned}
\end{equation}
The Cooper pairs, Bosons with zero spin, can be condensed to enter the superconducting state at Bose-Einstein condensation (BEC) temperature $T_{BEC}$, i.e., the superconducting transition temperature $T_c$, due to BCS-BEC crossover in high-$T_c$ cuprates~\cite{Supplemental,Chen-RevModPhys-2024,Chen-npj-2024,Zhu2025,Melo1993,Deutscher1999,Sous2023}. The transition from their excited states to the ground state releases energy, $\Delta E_{n0}=E_n-E_0$, closely related to the superconducting gap $E_\mathrm{sc}$ due to BEC~\cite{Mourachkine2002} (Fig.~\ref{fig:3}). We thus have,
\begin{equation}
\label{eqn:4}
\begin{aligned}
    T_c = T_{BEC},
    \end{aligned}
\end{equation}
\begin{equation}
\label{eqn:5}
\begin{aligned}
    E_{n,sc} = \Delta E_{n0} = E_n - E_0,\quad (n = 1, 2, 3 \dots),
    \end{aligned}
\end{equation}
where $n$ represents the number of the superconducting gap. The eV-scale ionic bonds, approximately 2 orders of magnitude larger than antiferromagnetic or electron-phonon coupling (Fig.~\ref{fig:1}(b)), ensure that Cooper pairs can be formed above $T_c$ or even at room temperature, as confirmed by Figs.~3.33, 3.34 and section~3.13 of chapter~3~\cite{Mourachkine2002} and Refs.~\cite{Gomes2007,Kondo2011}. The BEC, especially the potential room-temperature BEC mechanism in high-$T_c$ ionic oxides, will be the most important challenge and open-ended question~\cite{Shi2025}.

The \textbf{\ce{e^-}-O-\ce{e^-}} picture shown in Fig.~\ref{fig:2}(b) is a breakthrough insight into the electron-pairing mechanism in high-$T_c$ oxides and resolves a 40-year puzzle from the physical essence of eV-scale ionic bonds (the strongest and dominant interaction), oxygen-electron interaction and the large two-electron ionization energy of metal atoms, completely different from several mainstream theoretical images~\cite{Mourachkine2002,Navinder2021}. The strong ionic bonds ``push'' electrons towards oxygen atoms and the oxygen nuclei strongly attract electrons within a short range, which lead to an indirect eV-scale mutual ``attraction'' between the two electrons and a large pseudogap E$_{pg}$ (Fig.~\ref{fig:3}), required to form a Cooper pair~\cite{Mourachkine2002,Gomes2007,Kondo2011}. At the same time, electron pairs form around oxygen atoms, leading to small electron-pair size, for example, $\xi_c$$\sim$15~$\textrm{\AA}$ and $\xi_{ab}$$\sim$70-80~$\textrm{\AA}$ in electron-doped \ce{Nd_{2-x}Ce_xCuO4}, much smaller than the BCS electron-pair size about 400-10$^4$~$\textrm{\AA}$~\cite{Mourachkine2002}. Similarly, for the hole-doped cuprates and nickelates, the eV-scale ionic bonds overcome the repulsion of metal cations to holes (h$^+$), driving the holes with $d$-wave symmetry to concentrate towards metal cations to increase their valence states (Fig.~\ref{fig:2}(c)), eventually reaching equilibrium and forming an itinerant M-bridged \textbf{\ce{h^+}-M-\ce{h^+}} (M=Cu, Ni, etc.) hole pairing picture (Fig.~\ref{fig:2}(d)). It has smaller hole-pair size about 1-15~$\textrm{\AA}$ because ‌hole wavefunctions exhibit significantly poorer spatial extension compared to electron wavefunctions~\cite{Mourachkine2002}.

\begin{figure}
    \centering  \includegraphics[width=\linewidth]{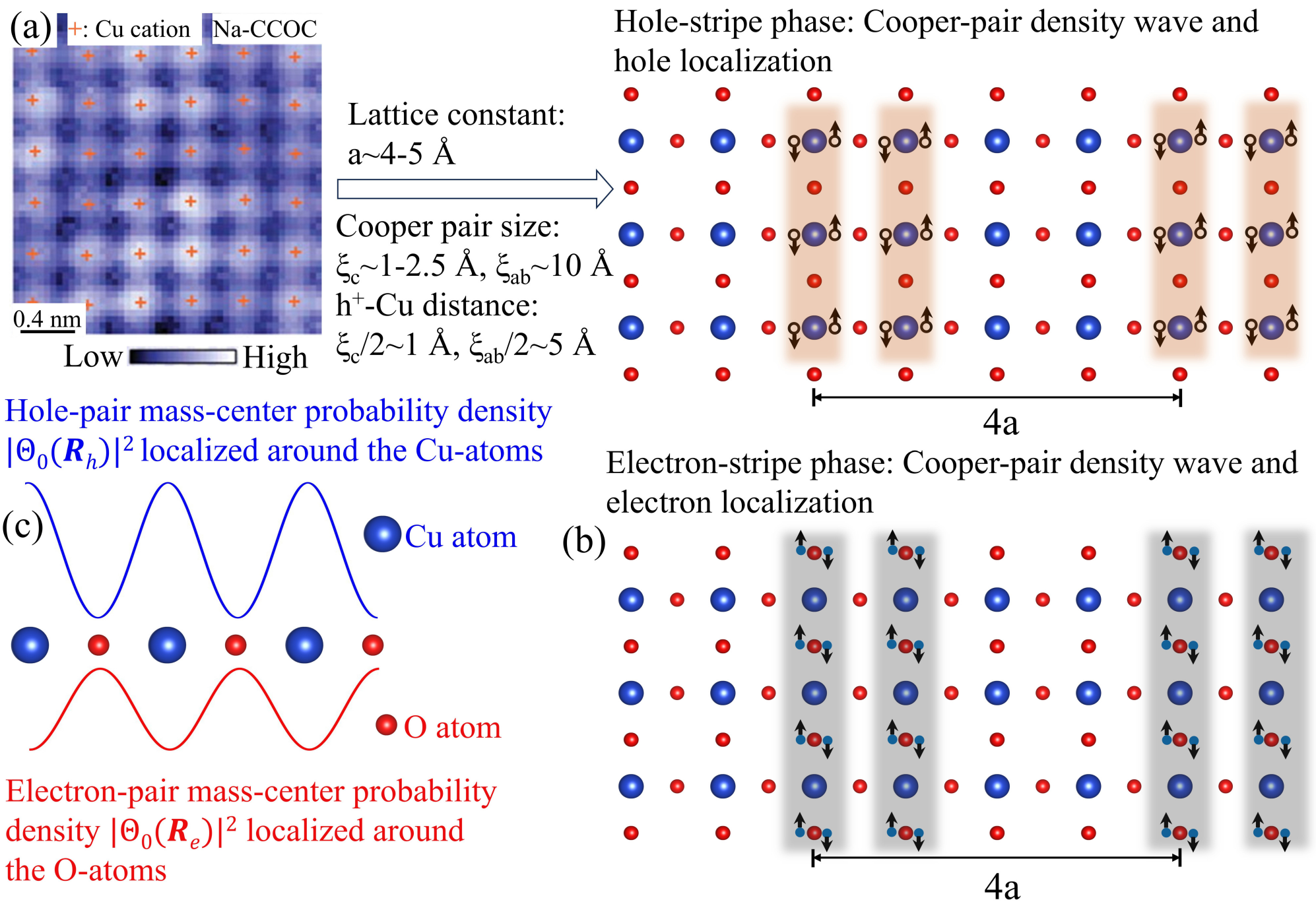}
    \caption{Cu-bridged hole and O-bridged electron pairing \textbf{\ce{h^+}-Cu-\ce{h^+}} and \textbf{\ce{e^-}-O-\ce{e^-}}, and their charge-stripe phases. (a) The STM constant-current topographic image of Na-CCOC~\cite{Kohsaka2007}, an irrefutable evidence of our \textbf{\ce{h^+}-Cu-\ce{h^+}} picture. It indicates that holes highly gather around Cu ($+$) cations and the 4$a$-period hole-stripe phase is formed~\cite{Vojta2008}, as Cooper-pair density wave and hole localization states. (b) The 4$a$-period O-bridged \textbf{\ce{e^-}-O-\ce{e^-}} stripe phase in the \ce{CuO2} plane. (c) Probability density distribution of the center-of-mass of the hole and electron pairs along the -Cu-O-Cu- chain. The subscripts $h$ and $e$ in $\Theta_0 (\bm{R})$ represent holes and electrons, respectively.}
    \label{fig:4}
\end{figure}

The scanning tunneling microscope (STM) constant-current topographic image (Fig.~\ref{fig:4}(a)) of \ce{Ca_{1.88}Na_{0.12}CuO2Cl2} (Na-CCOC, $T_c$$\sim$21~K) clearly indicates that the holes closely gather around Cu cations and the 4$a$-period hole-stripe phase is formed along the -Cu-O-Cu- chain of the \ce{CuO2} plane~\cite{Kohsaka2007,Vojta2008}. Given the small size of the hole Cooper pairs ($\xi_c$=1-2.5~$\textrm{\AA}$, $\xi_{ab}$$\sim$10~\textrm{\AA})~\cite{Mourachkine2002} and the superconductivity of Na-CCOC, the holes must be paired. In the present case, the only pairing pathway is hole pairing with the Cu atom as a bridge, powerfully confirming our Cu-bridged $d$-wave small-sized hole pairing picture: \textbf{\ce{h^+}-Cu-\ce{h^+}} (Fig.~\ref{fig:2}(d)). Similarly, as Cooper-pair density wave and electron localization states, the 4$a$-period electron-pair stripe phase in the \ce{CuO2} plane with O-bridged small-sized electron pairs \textbf{\ce{e^-}-O-\ce{e^-}} is shown in Fig.~\ref{fig:4}(b). Moreover, Fig.~\ref{fig:4}(c) shows the probability density distribution of the center-of-mass of the hole (electron) pair along the -Cu-O-Cu- chain.

Considering the similarity of cuprates and nickelates, our \textbf{\ce{e^-}-O-\ce{e^-}} and \textbf{\ce{h^+}-M-\ce{h^+}} images (Fig.~\ref{fig:2}) are also applicable to the electron-doped and hole-doped nickelates, respectively. In fact, our carrier-pairing pictures are determined by the physical essence of strong ionic bonding in crystals. In addition to high-$T_c$ cuprates and nickelates, they are also applicable to all ionic superconductors including iron-based~\cite{Hideo2018} and other new prospective superconductors, in which anions are responsible for electron pairing, while cations dominate hole pairing.

\textit{Conclusion}---We have made a breakthrough and developed a new theoretical framework for high-$T_c$ superconductivity, which resolves a 40-year puzzle of the microscopic carrier-pairing mechanism in complex cuprates and nickelates, formed by combining oxygen anions and metal cations together through ionic bonds. As we know, the crystal and electronic structure, the physical and chemical properties are totally determined by the chemical bonds, the foundation and essence, in a crystal. Similarly, the carrier-pairing mechanisms, as a change of the electronic structure, should also be intrinsically linked to the bonding nature of superconductors. We thus provide the missing link between ionic bonding and superconductivity. Generally, there are two different pairing mechanisms in the common superconductors. Different from the BCS electron-phonon-electron (\textbf{\ce{e^-}-Ph-\ce{e^-}}) pairing in charge of the superconductivity of metallic bonding superconductors, the high-$T_c$ oxides exhibit a totally new eV-scale ionic-bond-driven atom-bridged pairing picture \textbf{\ce{e^-}-O-\ce{e^-}} (\textbf{\ce{h^+}-M-\ce{h^+}}) with the strongest pairing strength and typical physical features (Eqs.~(\ref{eqn:1})-(\ref{eqn:3})).

Our pairing pictures root in the unshakable physical foundation, i.e., the affinity of O$^-$ (1.46~eV) and O$^{2-}$ (-8.08~eV), the large two-electron ionization energy of metal atoms ($\sim$15-28~eV, the low-valent cations), and the dominance of eV-scale ionic bonding (the strongest interaction), which plays a pivotal role for high-$T_c$ superconductivity in oxides. The other sub-eV and covalent-binding pairing mechanisms (see EM) are suspicious. The validity and universality of our itinerant \textbf{\ce{e^-}-O-\ce{e^-}} (\textbf{\ce{h^+}-M-\ce{h^+}}) pairing picture have been verified by 32 diverse experimental evidences (see EM and SM~\cite{Supplemental}), especially by the STM constant-current image combining with the small Cooper-pair size (Fig.~\ref{fig:4}(a)), serving as an irrefutable proof of our pairing mechanism. It is this ionic-bond-driven pairing image that dominates the unconventional high-$T_c$ superconductivity of oxide ceramics. In addition to the high-$T_c$ oxides, our new pairing mechanism, tightly grasping the physical essence of the ionic bonding, can naturally be extended to iron-based or other potential ionic-bonded superconductors. Our universal carrier-pairing picture also opens a new avenue for understanding high-$T_c$ mechanism and points out a new direction in searching for novel high-$T_c$ even room-temperature superconductors with robust ionic bonding at ambient pressure.

We further know from the \textbf{\ce{e^-}-O-\ce{e^-}} (\textbf{\ce{h^+}-M-\ce{h^+}}) picture that it is feasible to achieve room-temperature Cooper pairing in ionic oxides. If the issue of the room-temperature coherent condensation of Cooper pairs can be solved, the room-temperature superconductivity is hopeful to be achieved finally in oxides. According to textbooks~\cite{Atkins2010,Kittel2005,Sakurai1994}, we strongly believe that our transformative idea of \textbf{\ce{e^-}-O-\ce{e^-}} (\textbf{\ce{h^+}-M-\ce{h^+}}) electron (hole) pairing can arouse the widespread interest of superconducting scientists, especially theoretical physicists who are committed to developing high-$T_c$ theory. We hope to develop new theories of high-$T_c$ superconductivity rooted in our universal atom-bridged carrier-pairing image and BEC, and discover new high-$T_c$ even room-temperature superconductors with excellent superconducting properties, so as to further promote superconductivity research and benefit all mankind.

\textit{Acknowledgments}---We thank Professor Z.-M. Liao for fruitful discussion, manuscript review and editing; Y. Wang, A. Zaccone, J. Du, and B. Song for fruitful discussions. This work is supported by the National Key Research and Development Program (2023YFB4604400).

%\textit{Acknowledgments}---We thank Professor Z.-M. Liao (Peking University) for fruitful discussion, manuscript review and editing; Y. Wang (Peking University), A. Zaccone (University of Milan), J. Du (Beijing University of Technology), and H. Li (Nanjing University) for fruitful discussions. This work is supported by the National Key Research and Development Program (2023YFB4604400).

%\textit{Author contributions}---J.-J.S. carried out the research, created the e$^-$-O-e$^-$ and h$^+$-M-h$^+$ pairing pictures and wrote the paper. Y.-H.Z. provided energy levels of neutral oxygen atom and its anions and participated in discussion.

\textit{Data availability}---The data supporting this study’s findings are available within the Letter.

$^*$Corresponding author:
jjshi@pku.edu.cn
%\nocite{*}

\bibliography{prl-shi}% Produces the bibliography via BibTeX.

\subsection{End Matter}
\begin{figure}
    \centering  \includegraphics[width=6 cm]{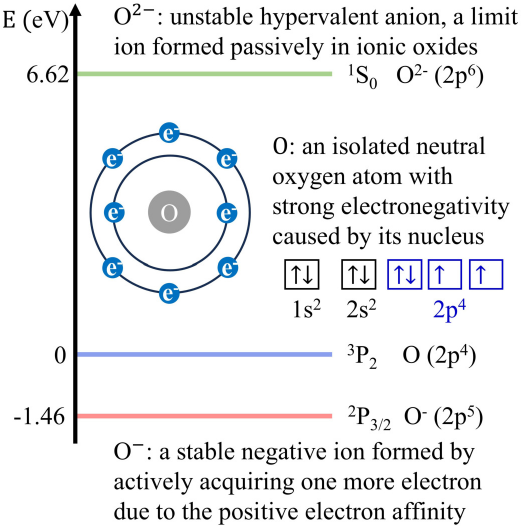}
    \caption{Energy levels of neutral oxygen atom O and its anions O$^-$ and O$^{2-}$~\cite{Atkins2010}.}
    \label{fig:5}
\end{figure}

\textit{O$^-$ repelling electrons and O$^{2-}$ unstable}---If using \ce{e-} to represent an electron, for an isolated oxygen atom O, we know~\cite{Atkins2010},
\begin{equation}
\label{eqn:6}
\begin{aligned}
     \ce{O + e^- -> O^- + 1.46 eV},
    \end{aligned}
\end{equation}
\begin{equation}
\label{eqn:7}
\begin{aligned}
    \ce{O^- + e^- -> O^{2-} + (-8.08) eV}.
    \end{aligned}
\end{equation}
Equation~(\ref{eqn:6}) demonstrates that the strong attraction between the atomic nucleus and electrons enables a neutral oxygen atom to readily gain an extra electron while releasing 1.46~eV of energy, forming an O$^-$ anion. In contrast, Eq.~(\ref{eqn:7}) shows that the O$^-$ anion repels additional electrons. In fact, adding another electron to form O$^{2-}$ requires an energy input of 8.08~eV, indicating its instability. Combining these results, we have,
\begin{equation}
\label{eqn:8}
\begin{aligned}
    \ce{O + 2e^- -> O^{2-} + (-6.62) eV}.
    \end{aligned}
\end{equation}
This equation reveals that the overall transition from a neutral O atom to an \ce{O^{2-}} anion is an endothermic process requiring 6.62~eV. Thus, while a neutral O atom can easily capture one electron to form the stable \ce{O-} anion due to the strong nuclear attraction, the subsequent electron repulsion in \ce{O-} outweighs this attraction, rendering the formation of a stable \ce{O^{2-}} anion unfavorable (Fig.~\ref{fig:5}).

\textit{Ionic bond or covalent bond?}---Let us now address the fundamental question: are cuprates and nickelates bonded via ionic or covalent bonds, and which one predominates? Considering the complexity of the crystal structure and the constituent atoms with more than two elements, we estimated the atomic weighted ionicity, i.e., fractional ionic character, of bonds in cuprates and nickelates from the simple Pauling’s ionicity~\cite{Phillips1973}, i.e., $f_i$=1-$\exp[-(X_A-X_B)^2/4]$ with $X_A$ and $X_B$ being the electronegativity {\color{blue}(see Fig.~S3 in SM~\cite{Supplemental})} of atoms A and B to form the bond. Although Pauling’s scale often underestimates ionicity, e.g., the calculated $f_i$$\approx$$71$\% much smaller than the well-recognized value ($f_i$>$90$\%) in \ce{NaCl}~\cite{Phillips1973}, it can still provide a quantitative reference.

\begin{table}[t]
\centering
  \caption{Ionicity of bonds in cuprates and nickelates.}
  \label{tbl:1}
  \begin{ruledtabular}
  \begin{tabular}{lclc}
    Cuprate & \makecell[c]{Ionicity}  & Nickelate & \makecell[c]{Ionicity}  \\
\colrule
  \ce{YBa2Cu3O7}  & 0.609  & \ce{LaNiO2}  & 0.594 \\
  \ce{Bi2Sr2CaCu2O8}  & 0.591  & \ce{PrNiO2}  & 0.590 \\
  \ce{Bi2Sr2Ca2Cu3O10}  & 0.596  & \ce{NdNiO2}  & 0.588 \\
  \ce{Tl2Ba2Ca2Cu3O10}  & 0.608  & \ce{La3Ni2O7}  & 0.625 \\
  \ce{HgBa2CuO4}  & 0.625  & \ce{La2PrNiO7}  & 0.668 \\
  \ce{HgBa2Ca2Cu3O10}  & 0.618  & \ce{La2PrNi2O7}  & 0.623 \\
  \end{tabular}
  \end{ruledtabular}
\end{table}

Table~\ref{tbl:1} shows that the ionicity for representative cuprates and nickelates ranges from 0.59 to 0.67, indicating ionic bonds dominate over covalent bonds. However, we note that the (Cu $3d_{x^2-y^2}$, O $2p_{x(y)}$) covalent bonds primarily determine the structural framework of the quasi-two-dimensional \ce{CuO2} planes~\cite{Xiang2022}, with two spin-antiparallel electrons constituting the covalent bond being attracted by the two nuclei to form a localized state between them. Importantly, semiconductor physics tells us that, in covalently bonded semiconductors such as Si and Ge, electrical conduction exclusively occurs through individual electrons~\cite{Kittel2005}, not electron pairs, questioning the conductive mechanism of electron pairs constituting covalent bonds. Moreover, for the most important hole-doped high-$T_c$ cuprates, two holes cannot form a covalent bond because of the strong Coulomb repulsion exerted by the two nuclei, indicating that efforts to seek a hole pairing mechanism based on covalent bonding are infeasible. Naturally, the ultimate resolution of the carrier pairing mechanism should be addressed within the framework of ionic bonding rather than covalent bonding. In fact, as the itinerant carriers, the electrons (holes) located at the indistinct boundary of the atom due to the ionic bond driving, furthest from the nucleus and not fully captured by it, that naturally contribute to conductivity as superconducting Cooper pairs (Fig.~\ref{fig:2}). This understanding applies equally to nickelates.

\textit{Experimental evidence}---Besides the $d$-wave Cooper pairs~\cite{Xiang2022,Stewart2017,Coleman1998}, the correctness of our \textbf{\ce{e^-}-O-\ce{e^-}} (\textbf{\ce{h^+}-M-\ce{h^+}}) image has been confirmed by the CuO$_2$-plane superconductivity in cuprates~\cite{Mourachkine2002}. Both the source of conducting holes and ‌the physical reason why they are confined to the ultra-thin CuO$_2$ planes (1-2.5 $\textrm{\AA}$) have also been clarified (Evidence~2~\cite{Supplemental}). The structural complexity of high-$T_c$ oxides makes many oxygen atoms (electron-gathering centers) and metal atoms (hole-gathering centers) non-equivalent, directly causing spatial inhomogeneity of Cooper pairs, resulting in charge-stripe order and strongly anisotropic superconductivity~\cite{Mourachkine2002,Krakauer1988,Stewart2017}. More particularly, as the density wave of Cooper pairs and the localized state of holes, the 4$a$-period hole-stripe phase along the -Cu-O-Cu- chain within the CuO$_2$ plane observed in STM measurements (Fig.~\ref{fig:4}(a)), combining with the small Cooper-pair size~\cite{Mourachkine2002}, provides the most intuitive and powerful experimental evidence, i.e., an ironclad proof, for our $d$-wave \textbf{ h$^+$-Cu-h$^+$} small-sized hole pairing picture (Fig.~\ref{fig:2}(d))~\cite{Kohsaka2007,Vojta2008,Andrej2016,Marcel2004,Hamidian2016,Mazumdar2018}. The O-bridged electron-stripe phase along the -Cu-O-Cu- chain within the CuO$_2$ plane in electron-doped cuprates provides the clearest and solid experimental evidence for our \textbf{e$^-$-O-e$^-$} picture~\cite{Mazumdar2018,Eduardo2015,Reichardt2018}.

The pseudogap (E$_{pg}$$\sim$80 meV in Bi2212), i.e., the net attraction arising from the eV-scale pairing energy of two holes forming a Cooper pair via ionic bonding to overcome their eV-scale Coulomb repulsion (Fig.~\ref{fig:3}), is larger than the superconducting gap (E$_{sc}$$\sim$40 meV in Bi2212)~\cite{Hüfner2008}, required for the Cooper-pair condensation~\cite{Mourachkine2002}, and also much larger than the E$_{sc}$ of BCS superconductors ($\sim$1-6 meV)~\cite{Carbotte1990}. It is the carrier pairing that leads to the reduction of resistance at pseudogap temperature $T^*$ above $T_c$ or even at room temperature (Evidences~9, 11~\cite{Supplemental})~\cite{Mourachkine2002,Gomes2007,Kondo2011}. Both E$_{pg}$ and $T^*$ decrease with increasing of carrier concentration due to weakening of ionic bond strength~\cite{Hüfner2008,Stewart2017}. The five key characteristics of the pseudogap have been clarified satisfactorily by our \textbf{e$^-$-O-e$^-$} (\textbf{h$^+$-M-h$^+$}) picture, confirming the correctness of our carrier-pairing mechanism. Therefore, the critical challenge of realizing the room-temperature superconductivity naturally becomes how to achieve Cooper-pair coherent condensation at room temperature under the condition of ensuring a certain carrier concentration, such as 10$^{21}$ cm$^{-3}$~\cite{Mourachkine2002,Wang1987}, which will be the most crucial challenge to be solved for realizing room-temperature superconductivity in ionic oxides at present~\cite{Shi2025}.

The two intrinsic energy gaps are caused by the formation and condensation of Cooper pairs in cuprates (Fig.~\ref{fig:3})~\cite{Mourachkine2002,Hüfner2008}, respectively. The $T_c$ and E$_{sc}$ of cuprates have a dome-like dependence on the carrier concentration and Cu-O spacing, closely related to the Cu valence state, non-copper cation concentration, and pressure~\cite{Mourachkine2002,Park1995,Scholtz1992}. The Cooper-pair size is as low as 1-15 $\textrm{\AA}$, leading to superconducting fluctuations~\cite{WangNingning2024,Mourachkine2002,Wang2022,Zhou2022,Inderhees1988,Pagnon1991}, much smaller than the BCS Cooper-pair size of 400-10$^4$ $\textrm{\AA}$~\cite{Mourachkine2002}. The inverted dome-like dependence of Cooper-pair size on the carrier concentration~\cite{Mourachkine2002} can be understood from the fact that Cooper pairs can occupy a larger space at low concentration, and the repulsion of unpaired electron (hole) to paired electrons (holes) at high concentration leads to an increase in the Cooper-pair size. The in-plane coherence lengths are larger than those along the $c$-axis due to larger in-plane O-O (Cu-Cu) distance than that along the $c$-axis in cuprates~\cite{Mourachkine2002}. Our \textbf{e$^-$-O-e$^-$} (\textbf{h$^+$-M-h$^+$}) picture, rooted in the solid foundation of strong ionic bonding, oxygen-electron interaction and large two-electron ionization energy of metal atoms (the low-valent cations), is strongly supported by the moderate carrier concentration ($\sim$10$^{21}$ cm$^{-3}$)~\cite{Mourachkine2002,Kittel2005,Krakauer1988,Mazumdar2018,Wang1987}, the small magnetic moment of Cu$^{y+}$  and O$^{x-}$ ions~\cite{Mazumdar2018,Vaknin1987,Thio1988,Suh1994,Nokelainen2020,Rajan2023,Forsyth1988,Bisogni2016,Shamblin2018}, the difference between nominal and actual carrier concentrations in cuprates~\cite{Mazumdar2018}, and the electron-to-hole transition with increasing of the Ce concentration as well as $T_c$ enhancement as rare-earth ionic radius increases in Re$_{2-x}$Ce$_x$CuO$_{4-\delta}$ (Re=Pr, Nd, etc.)~\cite{Mazumdar2018,Dagan2004,Armitage2002,Michio2016,Naito2002}. Slight O deficiency is beneficial for improving the superconductivity of $n$-type cuprates, which is a natural conclusion of our picture~\cite{Mazumdar2018,Kawashima1994}. The $T_c$ of cuprates sensitively depends on O concentration due to its close correlation with the carrier concentration~\cite{Mourachkine2002}. Nickelates~\cite{Wang2024,WangNingning2024,Zhou2022,Nomura2022,Cheng2024,Zhou2025} have similar structures and properties to cuprates, proving the universality of our \textbf{e$^-$-O-e$^-$} (\textbf{h$^+$-M-h$^+$}) picture. Please refer to SM~\cite{Supplemental} for detailed physical interpretations for the 32 experimental evidences listed above.

The above experimental evidences provide clear and powerful proofs of the correctness and universality of our \textbf{e$^-$-O-e$^-$} (\textbf{h$^+$-M-h$^+$}) carrier-pairing picture in  high-$T_c$ oxides. Our picture also clarifies many experimental puzzles~\cite{Mazumdar2018}, such as the carrier transition from electrons to holes if increasing Ce-concentration in Re$_{2-x}$Ce$_x$CuO$_{4-\delta}$ (Re=Pr, Nd, etc.), the small magnetic moment of Cu$^{y+}$ and O$^{x-}$ ions, the difference between the nominal and true carrier concentrations, increasing of $T_c$ with the increase of rare-earth ion radius in (Re,Ce)$_2$CuO$_4$, slight O deficiency benefit for superconductivity enhancement in \emph{n}-type cuprates, and the illusion of electrons gathering around Cu$^{y+}$ cations originated from the weaker localization of electrons than holes in cuprates~\cite{Mourachkine2002,Eduardo2015}, etc. However, the ionic bonding results in the poor ductility and pseudo-metallic superconducting behavior of oxides~\cite{Shi2024}. It is the exceptional strength of ionic bonding and attraction of oxygen nucleus to electrons in the eV-scale, surpassing both the electron-phonon~\cite{Yan2023,Lanzara2001} and antiferromagnetic coupling~\cite{Mourachkine2002} in the sub-eV scale, that leads to unconventional high-$T_c$ superconductivity with the strongest carrier-pairing in cuprates and nickelates.

\ifarXiv
    \foreach \x in {1,...,\numbersupplementpages}
    {
        \clearpage
        \includepdf[pages={\x}]{\supplementfilename}
    }
\fi

\end{document}